\documentclass[twocolumn,showpacs,preprintnumbers,amsmath,amssymb,pra]{revtex4}
\usepackage{epsfig}
\usepackage{graphics}
\usepackage{graphicx}% Include figure files

\usepackage{bm}% bold math

\newcommand{\be}{\begin{equation}}
\newcommand{\ee}{\end{equation}}
\newcommand{\bea}{\begin{eqnarray}}
\newcommand{\eea}{\end{eqnarray}}
\newcommand{\ket}[1]{| #1 \rangle}

\begin{document}

\title{Process tomography of field damping and measurement of Fock state lifetimes by quantum non-demolition photon
counting in a cavity}

\author{M. Brune$^1$}
\email{brune@lkb.ens.fr}
\author{J. Bernu$^1$}
\author{C. Guerlin$^1$}
\altaffiliation{Present address: ETH Zurich, CH-8093 Zurich Switzerland}
\author{S. Del\'eglise$^1$}
\author{C. Sayrin$^1$}
\author{S. Gleyzes$^1$}
\altaffiliation{Present address: IOTA, 91127 Palaiseau Cedex}
\author{S. Kuhr$^1$}
\altaffiliation{Present address: Johannes Gutenberg Universit\"{a}t, Institut
f\"{u}r Physik, Staudingerweg 7, D-55128 Mainz, Germany}
\author{I. Dotsenko$^{1,2}$}
\author{J. M.~Raimond$^1$}
\author{S. Haroche$^{1,2}$}
 \affiliation{$^1$Laboratoire Kastler Brossel, Ecole
Normale Sup\'erieure, CNRS, Universit\'e P. et M. Curie,
24 rue Lhomond, F-75231 Paris Cedex 05, France\\
$^2$Coll\`ege de France, 11 Place Marcelin Berthelot, F-75231 Paris Cedex 05,
France}

\date{\today}

\begin{abstract}
The relaxation of a quantum field stored in a high-$Q$ superconducting cavity
is monitored by non-resonant Rydberg atoms. The field, subjected to repetitive
quantum non-demolition (QND) photon counting, undergoes jumps between photon
number states. We select ensembles of field realizations evolving from a given
Fock state and reconstruct the subsequent evolution of their photon number
distributions. We realize in this way a tomography of the photon number
relaxation process yielding all the jump rates between Fock states. The damping
rates of the $n$ photon states ($0\leq n \leq 7$) are found to increase
linearly with $n$. The results are in excellent agreement with theory including
a small thermal contribution.
\end{abstract}

\pacs{
 03.65.Ta,  %Foundations of quantum mechanics; measurement theory
 42.50.Pq %Cavity quantum electrodynamics; micromasers
 }

\maketitle

The goal of quantum process tomography is to determine experimentally the
matrix elements of the super-operator describing the evolution of a quantum
system's density matrix \cite{chuang}. This information is acquired by
preparing a set of test states and monitoring their subsequent evolution. The
method has been applied so far to spins in NMR experiments \cite{chuang2}, to
solid state qubits \cite{Howard}, to vibrational states of atoms
\cite{Steinberg} and to quantum gate operations \cite{Obrien,Blatt}. We
describe here process tomography applied to the photon number distribution of a
relaxing field stored in a high-$Q$ superconducting cavity, in which Fock
states are used as test states.

Cavity field relaxation is described by a rate equation which, restricted to
the photon number distribution $P(n,t)$, is \cite{exploring}
\begin{equation}
\frac{dP(n,t)}{dt}= \sum_{n'} K_{n,n'}P(n',t).
\end{equation}
Quantum electrodynamics predicts $K_{n,n}= - \kappa[(1+n_b)n+n_b(n+1)]$,
$K_{n,n+1}=\kappa (1+n_b)(n+1)$, $K_{n,n-1}=\kappa n_b n$, all the other
coefficients being $0$. Here, $\kappa$ is the damping rate of the field energy
and $n_b$ the mean number of blackbody photons at temperature $T$. The time
constant $-1/K_{n,n}$ is the lifetime of the $n$-Fock state. In this Letter, we
report a complete experimental determination of the $K_{n,n'}$ coefficients.

Our experiment relies on a QND procedure \cite{QND90} to count the number $n$
of photons stored in a cavity. It is based on the measurement of
cavity-field-induced light-shifts on Rydberg atoms crossing the cavity one by
one. It projects the field onto Fock states with high fidelity. By detecting
long sequences of QND probe atoms along single field realizations, we follow
the field evolution and observe the jumps between Fock states due to cavity
damping \cite{lifedeath,collapse}. By analyzing a large ensemble of field
trajectories, we now partially reconstruct the super-operator describing the
field relaxation process in the cavity and measure the lifetimes of individual
Fock states which scale as $1/n$ at zero temperature \cite{lu}. This study
provides insights into the physics of these highly non-classical states whose
production by random \cite{lifedeath,collapse} or deterministic
\cite{CircuitQED1} processes has recently been demonstrated.

\begin{figure}
\begin{center}
\includegraphics[width=0.48\textwidth]{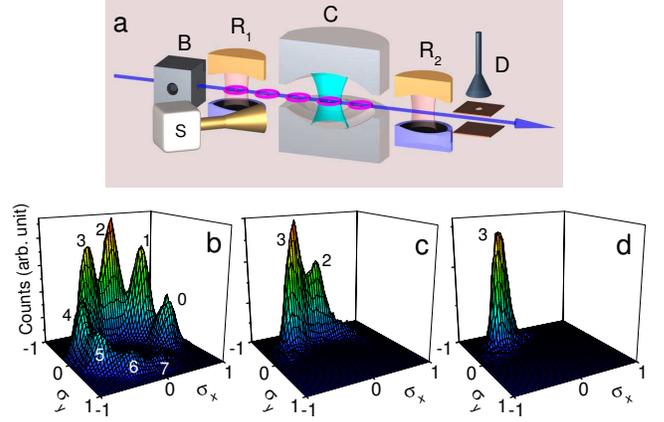}
\end{center}
\vspace{0 cm} \caption{\label{setup} (a) Scheme of experimental set-up. (b)
Histogram of the transverse atomic pseudo-spin after interaction with a
coherent field in $C$ (initial mean photon number 4.4). Photon numbers
associated to each peak are given. (c) Spin histogram after selection of the
$n=3$ Fock state. The $n=2$ peak is due to photon loss between selection and
measurement. (d) Same histogram as in (c) conditioned to a post-selection
measurement excluding events in which a photon is lost.} \vspace{-0.cm}
\end{figure}

Our setup \cite{lifedeath,rmp} is depicted on Fig.~\ref{setup}a. The high-$Q$
superconducting cavity $C$, operating at $51$ GHz, has a damping time
$T_c=1/\kappa=0.130$~s \cite{cavity}. A pulsed microwave source $S$, coupled by
diffraction on the mirrors' edges, can inject into $C$ a coherent field. The
cavity field is probed by a pulsed monokinetic stream ($v$=250~m/s) of Rubidium
atoms excited in box $B$ to the circular Rydberg state $g$ (principal quantum
number 50). Before $C$, the atoms experience in the low-$Q$ cavity $R_1$ a
$\pi/2$ pulse resonant on the transition to level $e$ (circular state with
principal quantum number 51). The atoms enter $C$ in the superposition
$(\ket{e}+\ket{g})/\sqrt{2}$. They undergo non-resonant light shifts in $C$,
resulting in a phase-shift $\Phi(n)$ of the atomic superposition which is, to
first order, linear in $n$. The phase-shift per photon is set to $\Phi_0\sim
\pi/4$.

We consider fields with a negligible probability of having $n>7$. In the Bloch
pseudo-spin representation, the atomic state at the exit of $C$ points along
one out of 8 directions equally distributed in the equatorial plane of the
Bloch sphere, corresponding to values of $n$ varying from $0$ (axis $Ox$) to
$7$. After leaving $C$, the atoms are submitted to a second $\pi/2$ pulse in
$R_2$ with an adjustable phase $\phi$ with respect to that of $R_1$. The atoms
are detected by the field ionization counter $D$ discriminating the states $e$
and $g$. Measuring the atomic energy after $R_2$ amounts to detecting the
atomic spin at the exit of $C$ in the direction at an angle $\phi$ with $Ox$ in
the equatorial plane of the Bloch sphere. On average, we detect one atom every
$0.24$ ms ($\sim500$ atoms detected during $T_c$).

Figure 1b shows a 3D histogram of the transverse atomic spin (components
$\sigma_x$ and $\sigma_y$) after interaction with a coherent field in $C$. Each
point in the Bloch sphere equatorial plane is obtained by measuring the average
value of spin projections, on a sample of 110 consecutive atoms crossing $C$ in
a $26$ ms time interval, much shorter than $T_c$. About 700 atoms are sent
across $C$, out of which we extract $\sim600$ atomic samples of 110 consecutive
atoms. The procedure is repeated 2000 times, yielding about $10^6$ spin
measurements. The histogram clearly shows that the direction of the atomic spin
is quantized.

After a measurement indicating a spin pointing towards a peak of this
histogram, the field is projected onto the corresponding Fock state. This is
checked by correlating two independent successive samples of $110$ atoms along
a single field realization. The first pins down $n$ and the second remeasures
it. Figure 1c shows the histogram of second measurements after selection of
$n=3$. It exhibits a main $n=3$ peak with an $n=2$ satellite due to field
relaxation during the $26$ ms time delay between the two measurements. This
satellite can be suppressed by post-selection. Figure 1d displays the histogram
of the intermediate results in sequences of $3$ measurements for which the
first and the last yield $n=3$. The single peak reveals that, at the
intermediate measurement time, the field contains exactly $3$ photons. We use
such single photon number peaks to calibrate the phase-shifts $\Phi(n)$.

Although the above method allows us to prepare Fock states and to observe
qualitatively their evolution, it lacks the resolution required for a precise
time analyzis. The interval between two measurements (26 ms) is longer than the
lifetime of $n=7$ (18 ms). We can however analyze the data in a more efficient
way, making better use of our atomic detection rate. On a single field
realization, each atom detected along direction $\phi$ provides one bit of
information $j$ ($j=0$ for $e$ and $j=1$ for $g$). After detecting $N$ atoms,
our knowledge of the field is described by an inferred photon number
probability distribution $p^i_N(n)$ linked to the initial distribution $P_0(n)$
by Bayes law: $p^i_N(n)= P_0(n)\Pi_N(n)/Z$ where $Z$ is a normalization and
$\Pi_N(n)$ is the product of $N$ functions $p(j,\phi|n)$, each describing the
information provided by one atomic detection: $p(j,\phi|n)=1/2[1+(-1)^j (A+B
\cos[\Phi(n)+\phi])]$ \cite{collapse,erratum}. For successive atoms, we use
four different values of $\phi$ (-1.74, -0.87, 0 and 0.54 rad) chosen so that
$p(j,\phi|n)$ is nearly maximal for $n=$6, 7, 0 or 1, respectively. The values
of $A$ and $B$, ideally 0 and 1, become -0.1 and 0.7, respectively, because of
experimental imperfections. For $N\sim100$, $p^i_N(n)$ converges to a Dirac
peak corresponding to the photon number given by the atomic spin analysis.

Let us call $P_N(n)$ the ensemble average $\langle p^i_N(n) \rangle$ over many
realizations in which the field is initially described by $P_0(n)$. As the
detection process is QND, we have $P_0(n)=P_N(n)$ for any $N$. In other terms,
$P_0(n)$ is a fixed point of the transform $P_0(n)\rightarrow \langle
P_0(n)\Pi_N(n)/Z\rangle$. This property allows us to determine $P_0(n)$ by
iteration of this transform starting with any initial non-vanishing
distribution, for instance the flat one $P_{\rm{fl}}(n)=1/8$. This method can
be applied for determining $P(n,t)$ at any time $t$ by selecting in each
sequence the sample of $N$ detected probe atoms starting at this time.

We first reconstruct in this way the evolution of $P(n,t)$ for a coherent field
injected by $S$ at $t=0$ and relaxing in $C$. A measurement sequence, involving
about 2750 atoms detected in $650$ ms, is repeated 2000 times. We reconstruct
$P(n,t)$ with the above procedure using $N=25$ atoms and 20 iterations. At each
time $t$, we start the iteration with $P_{\rm{fl}}(n)$. The temporal resolution
is $\sim6$ ms, much shorter than $T_c/7$.

\begin{figure}
\begin{center}
\includegraphics[width=0.43\textwidth]{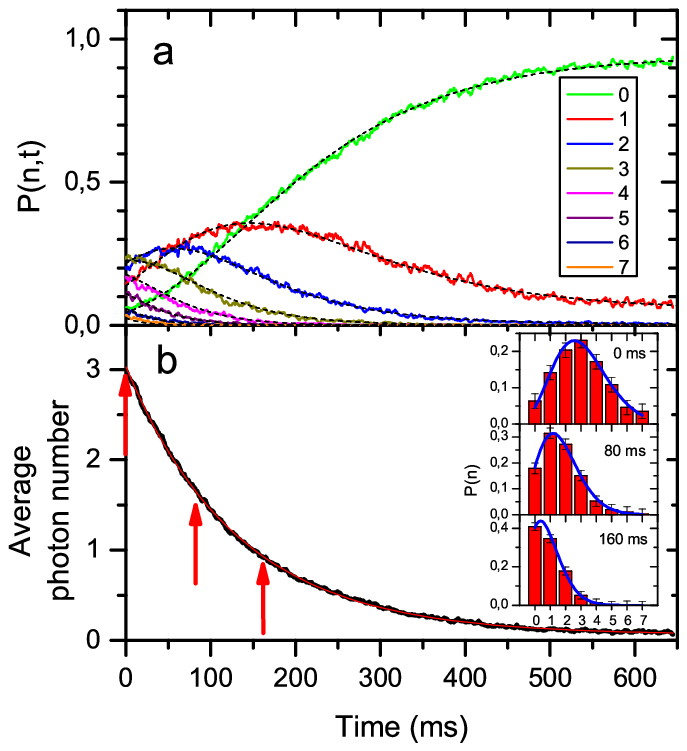}
\end{center}
\vspace{-0.5 cm} \caption{\label{fig2} Relaxation of a coherent state. (a)
Evolution of the photon number probabilities $P(n,t)$ ($n=0\ldots 7$ according
to the colour code defined in the inset). Black dotted lines are theoretical.
(b) Average photon number versus $t$ (solid black line) and exponential fit
(thin red line). Insets show the photon number distributions (red histogram)
and their Poisson fit (blue lines) at the three times shown by arrows.
}\vspace{-0.cm}
\end{figure}

Figure 2a presents $P(n,t)$ versus $t$ for $n=0$ to $7$ and Fig.~2b the time
evolution of the average photon number $\langle n\rangle = \sum_n nP(n,t)$.
According to theory, $\langle n \rangle$ is exponentially damped towards an
offset corresponding to the blackbody background. The experimental decay (solid
line) is indistinguishable from an exponential fit (red line), which yields
$T_c=132$~ms in agreement with the independently determined value of the cavity
damping time. The offset yields $n_b=0.06$, close to the theoretical value
(0.05) of the blackbody field at the cavity temperature, 0.8 K. The insets in
Fig.~2b present snapshots of $P(n,t)$ at three different times with the
corresponding Poisson fits. These histograms show, as theoretically predicted
\cite{exploring}, that the field remains coherent under the effect of damping
(at the limit where blackbody effects are negligible). The dotted line in
Fig.~2a presents a numerical solution of the theoretical rate equation using
the above determined values for $T_c$, $n_b$ and $\langle n \rangle$ at $t=0$.
It is in excellent agreement with the data (solid lines).

We go now one step further. By monitoring the decay of selected Fock states, we
determine the $K_{n,n'}$ coefficients without any a priori assumption about
their values. The same experimental data is processed in two steps. First, we
analyse separately the 2000 realizations of the experiment in order to select
individual Fock states. For each sequence, we compute, after each atom
detection, the new inferred photon number distribution $p^i(n,t)$ according to
Bayes law \cite{collapse}. We start the analyzis of each sequence with the
Poisson distribution determined above. Between atoms, we evolve the estimated
$p^i(n,t)$ according to the theoretical rate equation. This method gives, at
each time, the best estimate of the actual photon number distribution in each
realization. Except around quantum jumps, $p^i(n,t)$ is generaly peaked at a
single photon number value $n_0$. Whenever $p^i(n_0,t)> 0.7$ we assume that,
within a good approximation, the $n_0$ Fock state is present in $C$ at this
time, which we take as origin ($t=0$) for subsequent analysis of this Fock
state decay.

In a second step, we gather all atomic data following the selection of a given
$n_0$, obtaining thus ensembles of Fock state-selected field realizations. We
apply to each ensemble the iterative analysis described above, reconstructing
for each value of $n_0$ the subsequent $P_{n_0}(n,t)$ distributions. As in the
case of a coherent state, we use $N=25$ atoms and 20 iterations starting with a
flat initial distribution. Let us stress that this reconstruction procedure
does not rely on any theoretical assumption about the form of the relaxation
process. We made use of our theoretical knowledge of the $K_{n,n'}$
coefficients only in the first step of the data processing, in order to
optimize the selection of the initial Fock states.

\begin{figure}
\begin{center}
\includegraphics[width=0.48\textwidth]{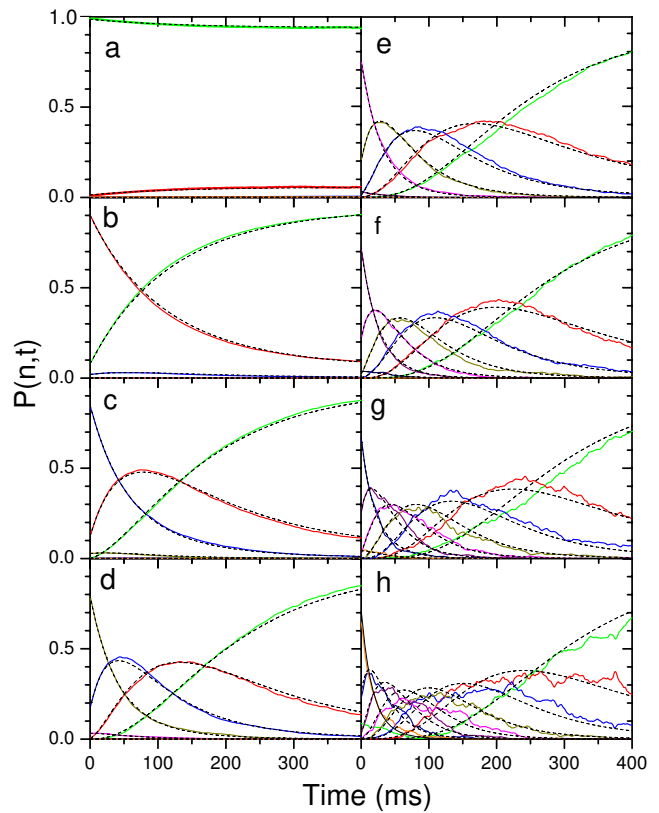}
\end{center}
\vspace{0 cm} \caption{\label{fig3} Relaxation of Fock states. (a) to (h)
Evolution of the photon number distributions $P_{n_0}(n,t)$ starting from the
 Fock states with $n_0=0\ldots 7$ respectively. Same color code as in Fig.~2a.
Dotted black lines are theoretical.} \vspace{-0.cm}
\end{figure}

Figure 3 shows in solid lines the reconstructed $P_{n_0}(n,t)$ distributions
versus time for $n_0=0$ to $7$ (a to h). In each frame, $P_{n_0}(n_0,t)$ is, as
expected, maximum at $t=0$, its value giving the fidelity of the Fock state
selection procedure. The other $P_{n_0}(n,0)$ values are small. At long time
(400 ms) the most probable photon number is always $n=0$, reflecting the
irreversible evolution of the field toward the thermal background close to
vacuum. For $n_0=0$ (Fig.~3a) $P_{0}(0,t)$ decreases slightly below 1, while
$P_{0}(1,t)$ reaches an equilibrium value close to $0.06$. This describes the
thermalization of the initially empty cavity. For $n_0=1$ (Fig.~3b) we observe
the exponential decay of $P_{1}(1,t)$, together with the increase of
$P_{1}(0,t)$, which describes the damping of a single photon into vacuum
\cite{lifedeath}. For $n_0>1$ (Fig.~3c to h), $P_{n_0}(n_0,t)$ decreases
exponentially at a rate increasing with $n_0$ (damping of the initial Fock
state). The $P_{n_0}(n,t)$ functions with $n=n_0-1,~n_0-2,~...,~1$ exhibit
bell-shaped variations. They peak successively, reflecting the cascade of the
photon number from $n_0$ down to vacuum.

In order to extract the damping coefficients, we fit the first $20$ ms of these
curves to a solution of Eq.~(1), leaving as free parameters the $K_{n,n'}$ and
the initial $P_{n_0}(n,0)$ values. The procedure is iterative. We get a first
approximation of $K_{n,n'}$ with $n$ and $n'\leq 1$ using the data of Fig.~3a
and b. We then determine the $K_{n,n'}$ with increasing indices by including
progressively in the fits the data of Fig.~3c to h, optimizing at each step the
previously determined parameters.

\begin{figure}
\begin{center}
\includegraphics[width=0.45\textwidth]{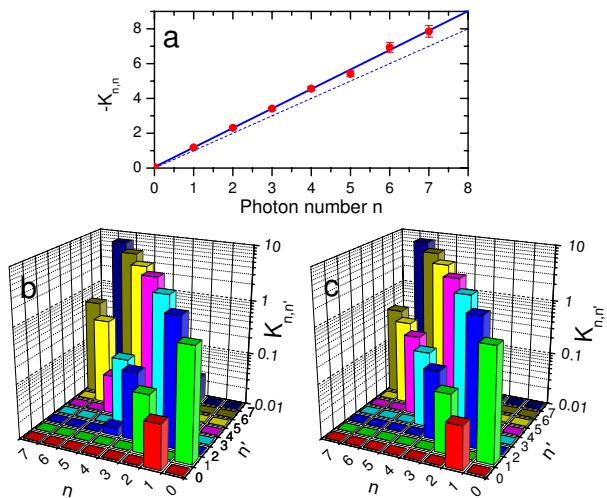}
\end{center}
\vspace{0 cm} \caption{\label{fig4} Measurement of the photon number
probability damping matrix elements $K_{n,n'}$. (a) Fock state damping rate
$-K_{n,n}$ versus $n$. Circles with error bars are experimental. The solid line
gives the theoretical values for $n_b=0.06$, the dotted line the expected rates
for $n_b=0$. (b) 3D plot of the measured non-diagonal elements $K_{n,n'}$ in
units of $\kappa$ (log scale). (c) Theoretical 3D plot of $K_{n,n'}$ ($n\neq
n'$) in units of $\kappa$ for $n_b=0.06$ (log scale).} \vspace{-0.cm}
\end{figure}

The obtained $-K_{n,n}$ values, which represent the decay rates of the $n$-Fock
states, are shown versus $n$ in Fig.~4a (in units of $\kappa$). As expected,
they vary linearly with $n$. The solid straight line corresponds to the theory
for $n_b=0.06$, while the dotted line shows the expected variation of $K_{n,n}$
at $T=0$ K. This constitutes the first measurement of Fock states lifetime for
$n>1$, exhibiting clearly the expected $1/n$ variation \cite{Martinis}.
Moreover, the departure of the experimental points from the dotted line shows
that our procedure is precise enough to be sensitive to the small effect of the
residual $n_b=0.06$ photon blackbody field on the lifetime of Fock states. The
non-diagonal $K_{n,n'}$ coefficients are shown (Fig.~4b) in a 3D plot, in
logarithmic scale. The big and small bars near the diagonal correspond to the
$K_{n,n+1}$ and $K_{n,n-1}$ coefficients, respectively. The latter, which
represent the thermal rates of photon upward jumps, are predicted to vanish for
$n_b=0$. The logarithmic scale is convenient to display together the
$K_{n,n+1}$ and $K_{n,n-1}$ coefficients which differ by about one order of
magnitude for $n_b=0.06$. All other non-diagonal coefficients are $0$ within
noise. Figure 4c shows for comparison the corresponding theoretical
coefficients for $n_b=0.06$.

The dotted lines in Fig.~3 are the result of a numerical integration of Eq.~(1)
using the values of $K_{n,n'}$ and $P_{n_0}(n_0,0)$ determined by our fit. The
excellent agreement with the experiment over the full $400$ ms time interval
shows the accuracy of our method.

This study demonstrates the power of QND photon number measurements to
investigate the quantum behavior of a field stored in a cavity. It directly
probes our theoretical understanding of field relaxation and clearly
illustrates the high sensitivity of large photon number states to decoherence,
their lifetime being (at $T=0$~K) inversely proportional to their energy. The
method is limited here to probing the rate equation for the photon number
probability distribution. In order to determine the full super-operator of
decoherence, we plan to monitor the evolution of the non-diagonal elements of
the field density operator in the Fock state basis, using a time resolved
quantum state reconstruction procedure demonstrated in Ref.~\cite{wignercat}.

{\bf Acknowledgements} This work was supported by the Agence Nationale pour la
Recherche (ANR), by the Japan Science and Technology Agency (JST), and by the
EU under the IP project SCALA. S.D. is funded by the D\'el\'egation
G\'en\'erale \`a l'Armement (DGA).


\begin{thebibliography}{99}
\bibitem{erratum}
Note that in [10]  $p(j,\phi|n)$ was misprinted. This error had no bearing on
the calculations presented there, which were performed with the correct
formula.
\end{thebibliography}
\end{document}